\documentclass[twocolumn,showkeys,superscriptaddress,10pt]{revtex4-1}
\usepackage{amssymb}
\usepackage{amsmath}

\usepackage{graphicx}
\usepackage{epstopdf}
\usepackage{colordvi}

\usepackage{amsmath,amssymb,graphicx}
\usepackage{epstopdf}
\newcommand{\beq}{\begin{equation}}
\newcommand{\eeq}{\end{equation}}
\newcommand{\beqa}{\begin{eqnarray}}
\newcommand{\eeqa}{\end{eqnarray}}

\begin{document}

\title{Fabry-Perot Filters with Tunable Josephson Junction Defects}

\author{Vincenzo Pierro}
\affiliation{Dept. of Engineering, University of Sannio, Corso Garibaldi, 107, I-82100 Benevento, Italy}

\author{Giovanni Filatrella}\email{Corresponding author: filatrella@unisannio.it}
\affiliation{Dept. of Sciences and Technologies, 
University of Sannio, Via Port'Arsa, 11, I-82100 Benevento, Italy}

\begin{abstract}We propose to take advantage of the properties of  long Josephson junctions to realize a frequency variable Fabry-Perot Filter that operates in the range $100-500$ $GHz$ with a bandwidth below $1$ $GHz$.
In fact, we show that it is possible to exploit the tunability of the effective impedance of the Josephson component, that is controlled by a dc bias, to tune, up to $10\%$ of the central  frequency, the resonance of the system. 
An analysis of the linearized system indicates the range of operation and the main characteristic parameters 
Numerical simulations of the full nonlinear Josephson element confirm the behavior expected from  the linear approximation.
\end{abstract}

\date{\today}
\keywords{Josephson devices; Superconducting infrared, submillimeter and millimeter wave detectors; Filters; Nonlinear waveguides \\
{\it To appear in Physica C}  \hfill DOI: {10.1016/j.physc.2015.07.010}}

\maketitle

\section{Introduction}
\label{introduction}
Layered structures  have been introduced in microwave and quasi-optical devices to produce filters with a narrow  transmission bandwidth \cite{Pozar} and as filter in the high frequency domain \cite{Macleod}.
Layered structures, periodic or aperiodic, have also been proposed for fundamental physics applications with sophisticated optics \cite{Principe15} and microwave applicators \cite{Chiadini14}.
A frequency band can be selected introducing, in the terminal points of the periodic structures, suitable defects;
the structures thus constitute a sort of Fabry-Perot filter  in the optical and quasioptical band that alternate periodically two different sections $H$ and $L$ of high and low distributed capacitance, respectively. 
Fabry-Perot like devices provide high precision measurements set-up for demanding applications \cite{Pitkin11,Addesso13}.
The defect ordinarily consists in a linear Transmission Line (TL), thus a well known (and commonly employed) system can be modelled as a periodic array of $\lambda/4$ TL that introduces identical delays, while the defect consists in still another TL of different length.
The periodic structure determines a band gap; if one introduces a defect, a narrow resonance appears in the bandwidth.
A natural limit of this technology is that the resonant frequency is determined by the physical characteristics of the defects; therefore, to tune the central frequency one should build (and insert) a different defect. 
Therefore, to overcome this complication superconducting tunable filters (with lumped parameters) have been explored  to realize electronically controlled devices without mechanical tuning \cite{Kaplunenko04,Zhou07,Mohebbi09}. 
In this context, we propose to employ as defects Josephson Junctions Transmission Lines (JJTL) \cite{Barone82,Likharev86,Jung14} to obtain a tunable filter in the sub-THz region. 
The motivation is twofold: Josephson Junctions (JJ) are very fast superconducting elements, suitable to build high frequency and low noise transmission lines \cite{Butz13,Chaudhuri14,Obrien14}, pseudocavities \cite{Zeuco13}, and generators \cite{Savelev06,Ozyuzer07,Delfanazari13,Pedersen14} capable to perform even near the THz region.
Moreover, JJ's are nonlinear element whose (effective) inductance can be tuned by an external dc bias \cite{Jung14,Hutter11}.
The latter property of JJ allows to control the effective inductance, at high frequency, with a change of the direct current through the superconducting element \cite{Pedersen73,Salehi07}.
In this work we examine how the main properties of such electronically controlled defect can be exploited to design a tuinable filter. 
The aim is to show that one can tune some properties of the device with a (relatively) simple change of the applied current. 
The main superconducting TL Josephson element consist of a so-called long JJ \cite{Barone82,Likharev86}. 
Design capability for JJ are highly developed, as demonstrated by the recent upsurge of superconducting metamaterials\cite{Maimistov10,Lapine14}, especially for quiet quantum measurements \cite{Jung14,Castellanos08}. 
It is therefore conceivable that the proposed concept can be considered for practical purposes.
The work is organized as follows. 
In Sect. \ref{model} we retrieve, in the $ABCD$ matrix formalism \cite{Frichey94}, the dependence of the JJTL as a function of the dc bias. 
In Sect. \ref{results} we show in the linear approximation  the main properties of the proposed structure: bandwidth and tunability. 
We also numerically investigate the full nonlinear JJTL and compare its behavior to the approximated linear analysis. 
Sect. \ref{conclusions} concludes.

\section{Model}
\label{model}

\begin{figure}[htbp]
\centerline{\includegraphics[width=1.\columnwidth]{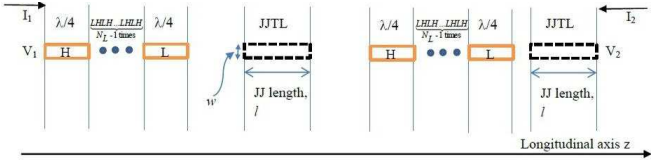}}
\caption{Schematic top view (not in scale) of the planar microstrip circuit representing  a chain of $\lambda/4$ standard TL (solid lines rectangles) joined to JJTL defects (dashed lines rectangles). The ac component travels along the z-axis, for the space between the sections is only shown to distinguish the various elements. A dc current bias is only fed through the JJ element. In the following  of the paper we take the number of TL sections $N_L=8$. }
\label{fig:TL}
\end{figure}

The constitutive equations for the JJTL element of Fig.~\ref{fig:TL} read \cite{Maimistov10}:
\begin{eqnarray}
& &\frac{\partial V}{\partial z} = -L_0\frac{\partial I}{\partial t}
\label{eq:VvsI}\\
& &\frac{\partial I}{\partial z} = 
- C_j \frac{\partial V}{\partial t} - I_c  \sin{\phi} + I_b - G_j V
\label{eq:JJ}\\
& &\frac{\hbar}{2e} \frac{\partial \phi}{\partial t} = V
\label{eq:JJV}
\end{eqnarray}
where $\hbar$ is the Planck constant, $e$ is the elementary charge, $V$ and $I$ are the distributed voltages and currents, respectively; $C_j$, $G_j$, and $I_c$ are the capacitance, conductance and  maximum (or critical) current per unit length of the superconductive element; $L_0$ is the geometric inductance per unit length.  Eqs.~(\ref{eq:VvsI},\ref{eq:JJ},\ref{eq:JJV}) amount to a sine-Gordon equation \cite{Barone82,Likharev86}.
The variable $\phi$ is the gauge invariant phase difference between the superconducting wave functions \cite{Barone82} that determines the Josephson supercurrent $I_c\sin(\phi)$, the essence of the Josephson effect, as it corresponds to the tunneling of Cooper's pairs. 
Finally, Eq.(\ref{eq:JJV}) is the phase-voltage Josephson relation.
From the standpoint of the electromagnetic propagation it corresponds to a nonlinear distributed  inductance \cite{Mohebbi09}:
\beq
L_j =\frac{\hbar}{2eI_c}\left[1-\left(\frac{I_b}{I_c}\right)^2 \right]^{-1/2}.
\label{eq:Lj}
\eeq
This is the key property of the Josephson effect that we want to exploit: the effective inductance $L_j$ can be tuned by the external dc bias $I_b$, in analogy to fluxometers (SQUIDs) or other microwave components \cite{Pedersen73}.

Linearizing the JJTL Eqs.~(\ref{eq:VvsI},\ref{eq:JJ},\ref{eq:JJV}) and using the distributed inductance (\ref{eq:Lj}) in the frequency domain we find:
\beqa
\frac{dV}{dz} &=& - j\omega L_0 I 
\label{JJTL1} \\
\frac{dI}{dz} &=& - \left( j\omega C_j + \frac{1}{j\omega L_j} + G_j \right) V \equiv  -j\omega C_{eq}V,
\label{JJTL2}
\eeqa
where $V$ and $I$ are the voltage and current phasors and we use the substitution rule $\partial (^.)/\partial t \rightarrow j\omega$. The factor $C_{eq}$ in Eq.(\ref{JJTL2}) amounts to an equivalent condenser $C_{eq}\equiv C_j-1/(\omega^2 L_j)+G_j/j\omega$, that depends on the dc bias $I_b$ through Eq.(\ref{eq:Lj}). 
Through the whole paper we use the term {\it frequency} as a synonymous of {\it angular velocity}.
The linear analysis of the TL gives the (secondary) parameters $\beta$ and $Z_c$:
\beqa
\beta = \omega\sqrt{L_0C_{eq}},
\label{eq:beta} \\
Z_c = \sqrt{\frac{L_0}{C_{eq}}}.
\label{eq:Zc}
\eeqa

In the $ABCD$ matrix formalism \cite{Frichey94} the parameters $\beta$ and $Z_c$ are fundamental to derive the input-output  matrix  relations. 
For instance, the parameters $\beta$ and $Z_c$ of Eqs.(\ref{eq:beta},\ref{eq:Zc}) can be employed to derive the input-output relations of a JJTL:
\beqa
 \left( \begin{array}{c}
V_1\\
I_1 \end{array} \right) =&
\left( \begin{array}{cc}
\cos\left(\beta l\right) &jZ_c \sin\left(\beta l\right) \\ 
jZ_c^{-1}\sin\left(\beta l\right) & \cos\left(\beta l\right) \\
\end{array} \right) &
\left( \begin{array}{c}
V_2\\
I_2 \end{array} \right).
\label{eq:ABCDJJ}
\eeqa
In the following we shall indicate the matrix of the above Eq.(\ref{eq:ABCDJJ}) with $\hat{D}$.

\begin{table}[h!]
  \caption{
We have chosen the main physical parameters of the JJTL component close to well established technology \cite{Ustinov93,Carapella00}(we refer to Fig. \ref{fig:TL}):
$l$ and $d$ are the length of the JJTL element and of the high capacitance nonsuperconductive TL,  respectively (the length of the low capacitance TL is obtained by the $\lambda/4$ constraint). $R_j$ is the JJ resistance, $J_c$ the Josephson critical current density,  $h$ and $w$ the thickness and the width, respectively, while $\epsilon_r$ is the relative dielectric constant of the insulating barrier. 
The Josephson physics \cite{Barone82,Likharev86} dictates the other parameters:
 $I_c R_j=\Delta$ ($\Delta$ is the superconducting gap) and $L_0=\mu_0 h / w$. 
$C_H$ and $C_L$ are the capacitance per unit length of the $H$ and $L$ lines, respectively. 
We also denote with $L_{ind}$ the inductance per unit length of both $H$ and $L$ nonsuperconductive TL.
The input voltage source amplitude is $V_0=50 \mu V$, series connected with a generator impedance $R_G=0.29\Omega$, while the final edge of the array is terminated by a real resistance $R_L=0.29 \Omega$.
These figures are but an example of a realistic configuration with acceptable performances. 
With these parameters the usual normalized Josephson parameters read: $\omega_j \simeq  100 GHz $,  the Josephson penetration depth is $\lambda_j \simeq  40 \mu m$, the normalized dissipation $\alpha \simeq 0.006$. 
}
  \begin{center}
    \begin{tabular}{ccccccccccc}
    \hline
$\#$ &     $l$          & d               &        $R_j$          &    $J_c$         & $h$       &   w            & $\epsilon_r$     & $C_H$   & $C_L$  &  $L_{ind}$   \\
        &   $\mu$m    &  $\mu$m   &  $\Omega$         &      A/cm$^2$  &        nm & $\mu$m     &                   &      nF/m & nF/m  &  nH/m   \\
   \hline
    \hline
$(a)$ &  $360$       &  $250$     &    $50$                &  $28$             & $5$         & $1$        &   $ 50\,\,$       &   $88.5$  & $44.3$  &  $12.6$    \\
$(b)$ &  $180$      & $100$      &    $50$                &  $28$             & $5$         & $1$        &   $ 50\,\,$       &    $88.5$ & $44.3$  &  $12.6$   \\
   \hline
        \end{tabular}
\label{table:physical}
    \end{center}
\end{table}
The defect described by Eq.(\ref{eq:ABCDJJ}) has the structure of an ordinary defect \cite{Pozar} that introduces an appropriated phase delay  of the signal between the input and the output points \cite{note}. 
This defect can be combined with standard input-output relations for $\lambda/4$ elements \cite{Frichey94}. If $L$ and $H$ are two transmission lines with different capacitances ($C_L$ and $C_H$ for the low and high pieces, respectively) and with the same inductance $L_{ind}$, one obtains a band pass behavior due to the periodic structure.
For the complete  system (including the JJTL defect) of Fig.\ref{fig:TL}, the transmission matrix reads:
\beqa
T &=& \,\,  \underbrace{LH LH ... LH LH}_\text{$N_L$ times} \,\, \hat{D} \,\,   \underbrace{LH LH ... LH LH}_\text{$N_L$ times} \,\, \hat{D} \,\,  \hat{R}_L   \nonumber \\
&=&  \left(LH \right)^{N_L} \hat{D}  \left(L H \right)^{N_L} \hat{D} \,\, \hat{R}_L
\label{eq:chain}
 \eeqa
that completely describes the linear behavior (here, a termination is described by the resistive matrix  $\hat{R}_L$) . 
If we define $T$ as the result of the matrix multiplication in Eq.(\ref{eq:chain})
\beq
T = \left( \begin{array}{cc}
A' &B' \\
C' & D' \\
\end{array} \right), 
\label{prime}
\eeq
the input impedance $Z_i$ reads:
\beq
Z_i = \frac{V_i}{I_i} = \frac{A'R_L + B'}{C'R_L+D'}.
\label{Zi}
\eeq
Finally, denoting with $R_G$ the impedance (purely resistive) of the feeding TL, the transmission coefficient $\Gamma_i$ reads:
\beq
\Gamma_i = \frac{Z_i-R_G}{Z_i+R_G}.
\label{eq:Gamma}
\eeq
Equation (\ref{eq:Gamma}), together with Eq.(\ref{eq:Lj}) that is contained in the matrix $\hat{D}$, states that the transmission coefficient exhibits a dip that depends upon the dc bias current $I_b$. This is the main result of this paper, illustrated in the next Section.

\begin{figure}[htbp]
\centerline{\includegraphics[width=.74\columnwidth]{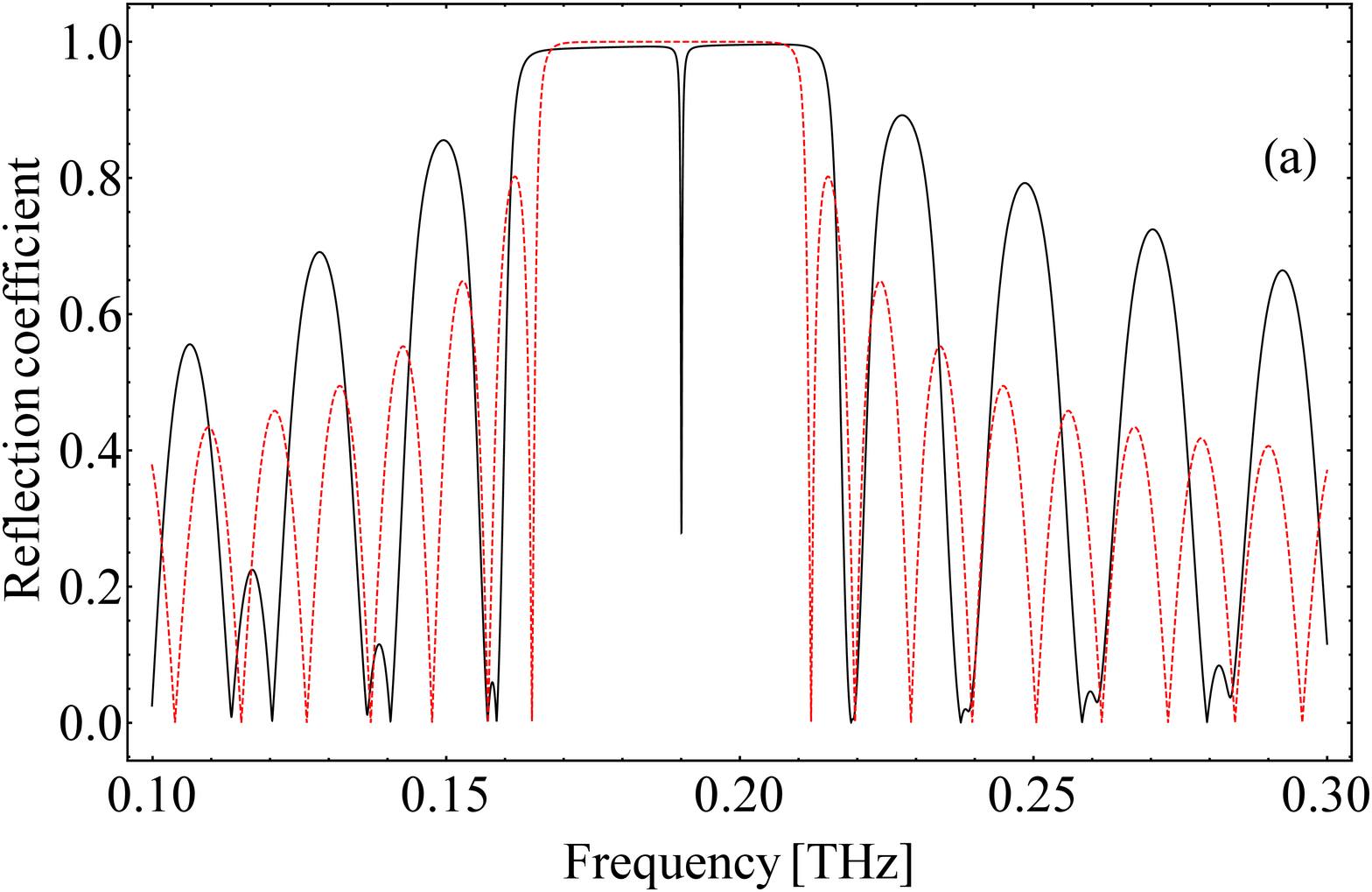}}
\vspace{0.5cm}
\centerline{\includegraphics[width=.74\columnwidth]{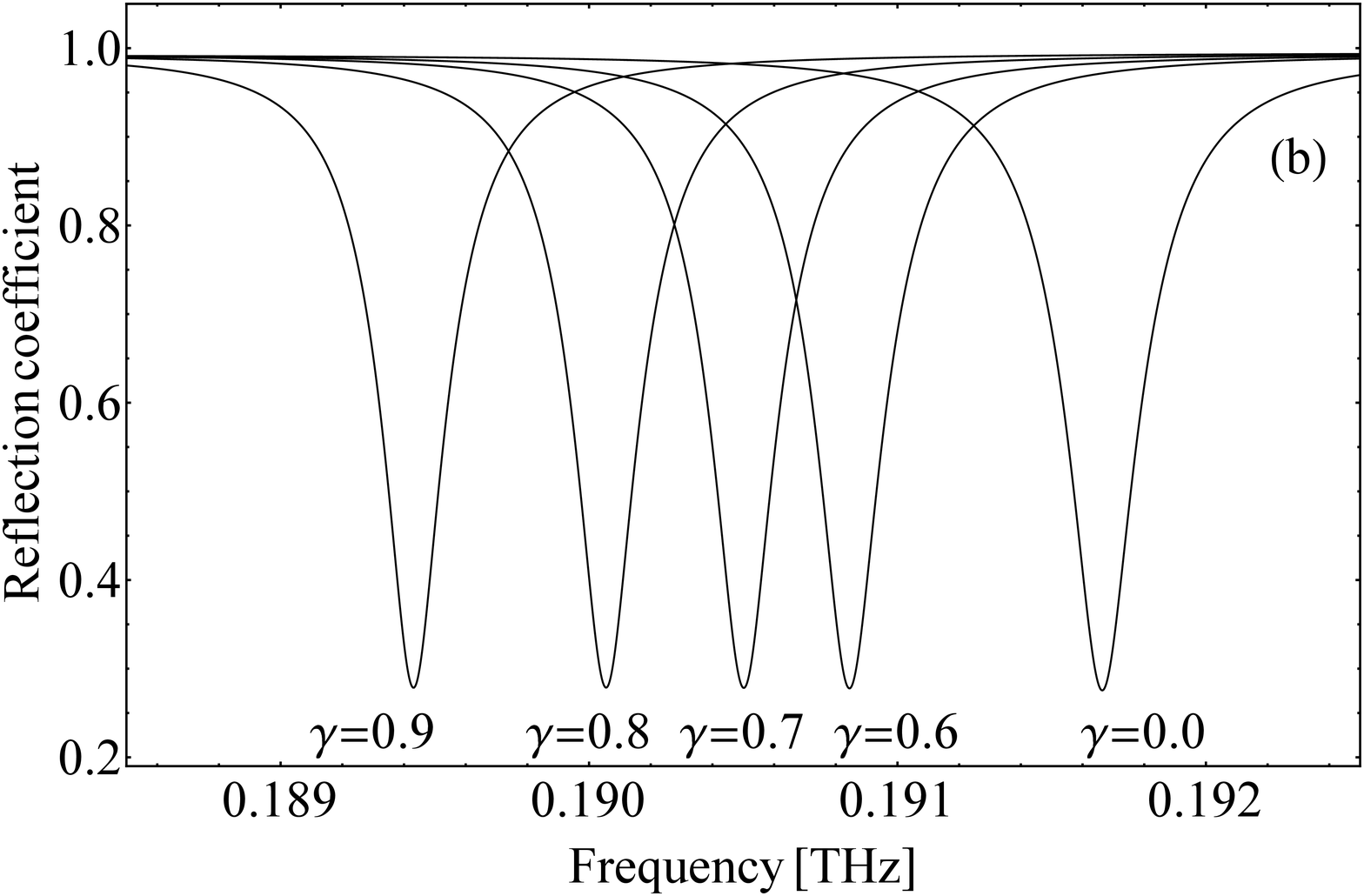}}
\caption{(color online) (a) Absolute value of the reflection coefficient $\Gamma_i$ as per Eq.(\ref{eq:Gamma}) for a TL interrupted by a JJ defects (black solid line) and of a TL without defect (red dashed line). 
The frequency, or more precisely the angular velocity, is measured in Hertz. 
The bias current reads $\gamma = I_b/I_c=0.8$.
(b) Absolute value of  the reflection coefficient  $\Gamma_i$ for different values of the current $\gamma$.
The other physical parameters are given in Table \ref{table:physical}.
}
\label{fig:reflection}
\end{figure}

\begin{figure}[htbp]
\centerline{\includegraphics[width=.85\columnwidth]{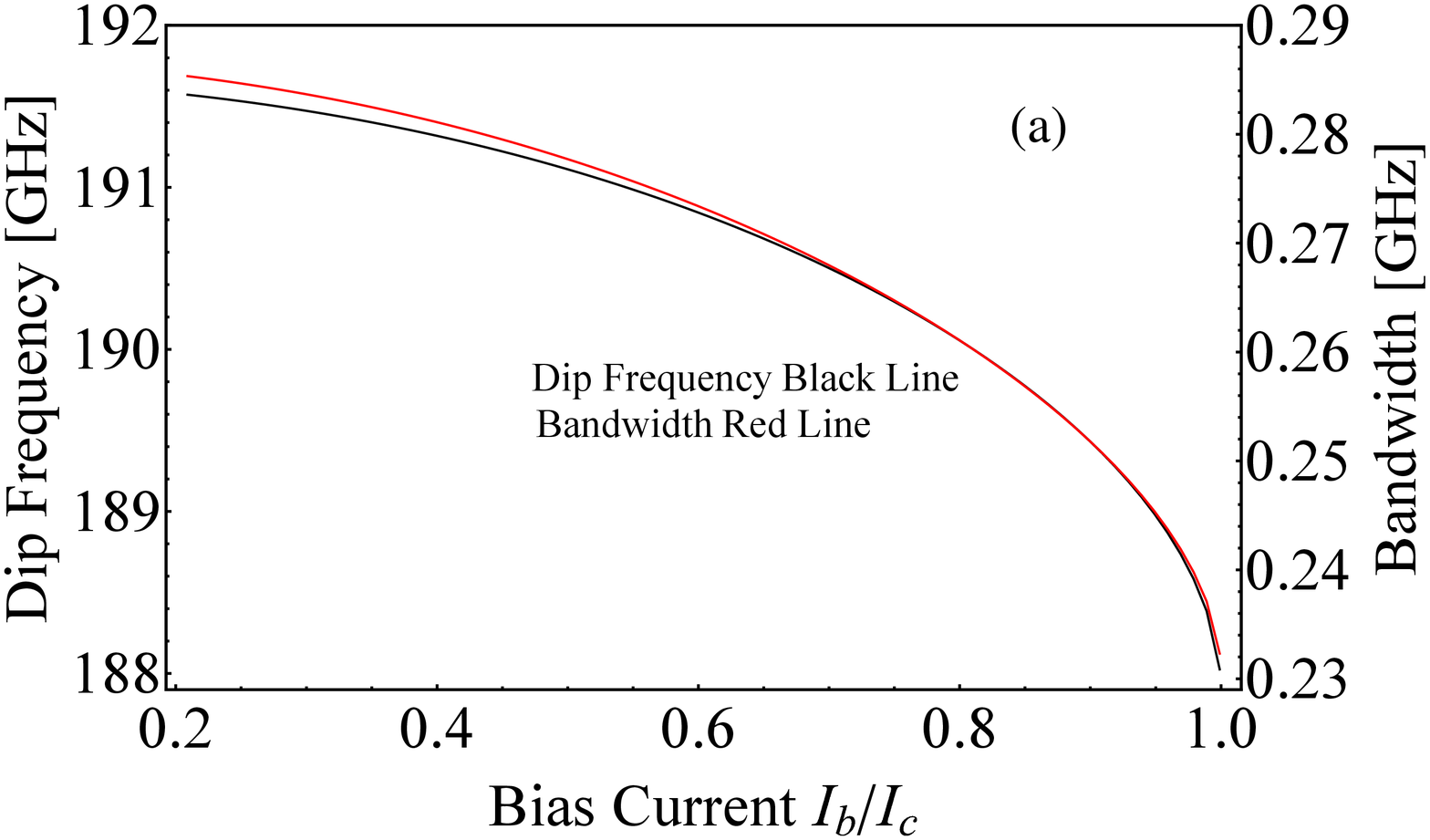}}
\centerline{\includegraphics[width=.85\columnwidth]{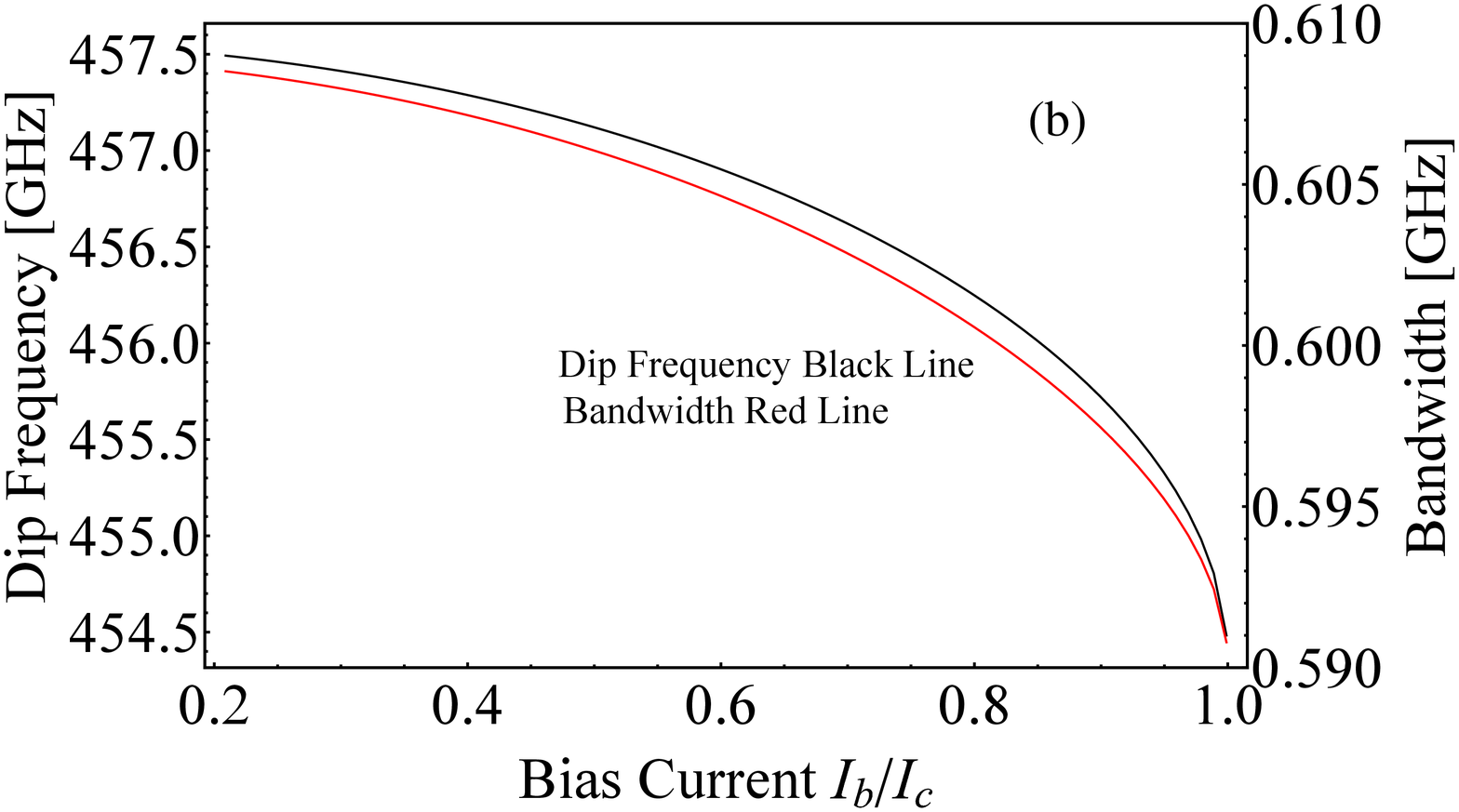}}
\caption{Effect of the bias current $\gamma=I_b/I_c$ (i.e., the bias current $I_b$ normalized to the Josephson junction critical current $I_c$) on two main parameters of the Fabry-Perot filter: the central frequency of the passing window and the linewidth of the band (see Fig. \ref{fig:reflection}). The physical parameters are given in Table \ref{table:physical} with the same letter.}
\label{fig:tunability}
\end{figure}
\section{Results}
\label{results}
To characterize the proposed Fabry-Perot filter, we include an input impedance and a load to obtain relevant quantities as the transmission coefficient. 
For simplicity, we consider a simple voltage source and a resistive load at the beginning and the end of the block diagram described in Sect. \ref{model}. 
For the parameters of Table \ref{table:physical} the reflection coefficient $|\Gamma_i|$ as a function of the frequency is displayed in Fig. \ref{fig:reflection}a at two values of the bias.
We stress that the resonance dip in Fig. \ref{fig:reflection}a can be tuned by a simple change of the bias current $\gamma = I_b/I_c$ in Eq.(\ref{eq:JJ}), as shown in Fig. \ref{fig:reflection}b. 
It appears that the central frequency can be tuned, within the allowed range of bias current, of as much as $3.5GHz$ (i.e., about $2\%$ of the resonance frequency, for this parameters choice).
This is shown in Fig. \ref{fig:tunability}:
the current $I_b$ can be tuned from $0$ (as negative value do not change the behavior of the device) to a maximum current that is the critical current of the Josephson element (above such current the junction develops a voltage and the propagation of the signal is not described by the present model). 
The Figure displays the effect of the bias for the two sets of parameters of Table \ref{table:physical}, to demonstrate the generality of the mechanism.

\begin{figure}[htbp]
\centerline{\includegraphics[width=.88\columnwidth]{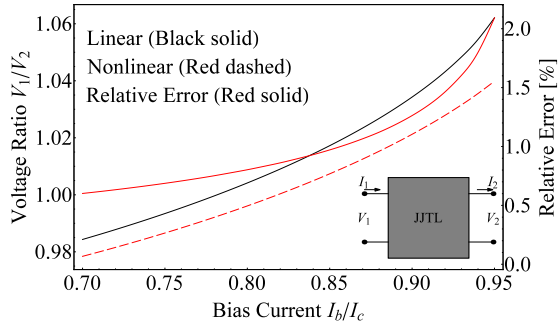}}
\caption{(Color online) Comparison of the input-output relation, see the inset, of the JJ transmission line (red, dashed) as per Eq.(\ref{eq:JJ}) with the linear prediction of the matrix $\hat{D}$ obtained through the approximation (\ref{eq:Lj}) (black, solid). 
The relative error of the linear approximation is displayed as red solid. 
The input voltage reads $V_0=0.05mV$. The other  physical parameters are given in Table \ref{table:physical}(a).}
\label{fig:linear}
\end{figure}

A key point for the validity of the approach of Sect. \ref{model}, and hence of the tunability described in Fig. \ref{fig:tunability}, is that the linearization of the inductance employed in Eq.(\ref{eq:Lj}) is valid. 
This has been checked for the values of Table \ref{table:physical}(a) in Fig. \ref{fig:linear}.
The comparison shows that the linear analysis correctly captures the main effect, as it reproduces the input-output voltage ratio trends. 
The discrepancy is about $0.5\%$ at low bias, to reach a maximum of about $1.5\%$ close to the critical current. 
This is to be expected, for the approximation (\ref{eq:Lj}) is based on a linearization of the sinusoidal  function in Eq.(\ref{eq:JJ}), that best performs at low bias.
It is important to underline that the comparison is performed without any fitting parameter. 
Also, the nonlinear model is sensitive to the amplitude of the input signal. In Fig. \ref{fig:linear} we have assumed an input signal of amplitude $V_0=0.05mV$. 

\begin{figure}[htbp]
\vspace{-0.0cm}
\centerline{\includegraphics[width=.75\columnwidth]{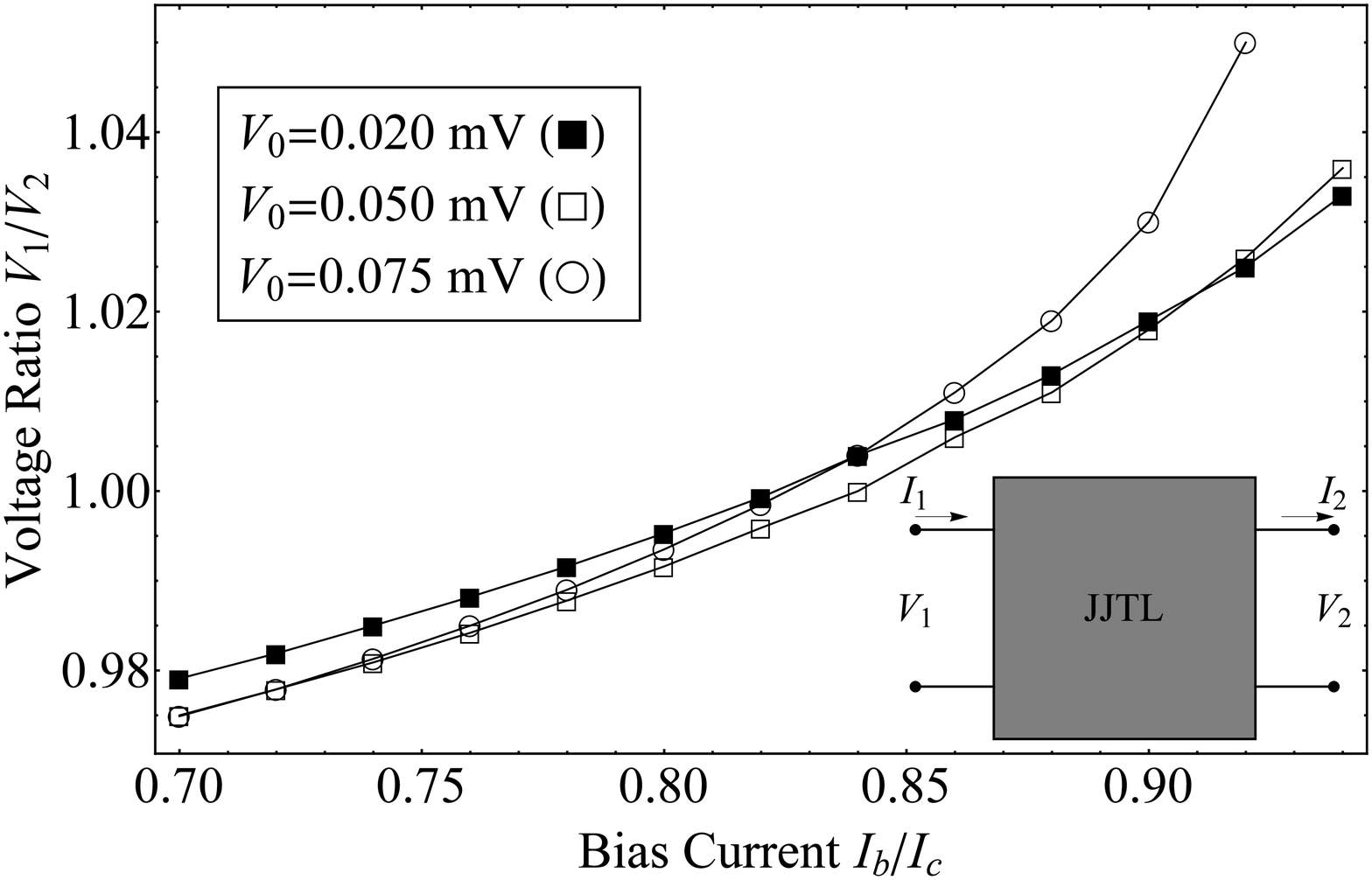}\hspace{0.64cm}}
\caption{The effect of the input voltage amplitude on the input-output relations, see the inset. The other  physical parameters are given in Table \ref{table:physical}(a).}
\vspace{-0.75cm}
\label{fig:inputV}
\end{figure}

The input voltage induces a current of the order $V_0/(R_G+Z_c)  \simeq 0.08 mA$; this value should be compared to the critical current of the junction $I_c l \simeq 0.1 mA$. We thus expect that the linear regime breaks up for larger values, see Fig. \ref{fig:inputV}. 

\vspace{0.6cm}

\section{Conclusions}
\label{conclusions}

We have shown that a superconducting metamaterial \cite{Lapine14} containing long Josephson junctions, inserted in a chain of $\lambda/4$ layers as a discontinuity, amounts to a tunable defect that moves the resonance characteristic frequency of a Fabry-Perot filter.
The change of the frequency, as in series JJ discrete TL \cite{Mohebbi09}, is achieved sweeping the dc current bias.
The effect can be regarded as relevant for applications in superconducting electronics and devices, for it allows to tune a Fabry-Perot filter with an externally dc bias.
Moreover, we have found that a linear analysis correctly captures the main effect (the nonlinear corrections are below $1.5\%$).
Some words of caution are in order, however. 
In the first place, we have overlooked fabrication limits and parameters tolerance, assuming that one can effectively produce the desired JJ. 
In the second place, we have fully examined the core part of the device, the nonlinear (superconducting) Josephson transmission line and the linear array, while a real device consists of other parts (e.g., control electronics to sweep the JJ bias current). 
Finally, the analysis has been performed with a single harmonic component (higher order expansions can be included as in Ref. \cite{Zeuco13}) and neglecting noise. The latter requires some attention, for the combination of noise and an ac term can induce undesired switches to finite dc voltage states \cite{Valenti14}.

\begin{thebibliography}{99}
\bibitem{note} A single small JJ, as a lumped element, cannot introduce this essential delay. 
However, a chian of coupled small JJcould work as well.
\end{thebibliography}
\end{document}